\documentstyle[12pt,aasms]{article}
\journalid{337}{15 January 1989}
\articleid{11}{14}
\begin{document}

\title{TGRS OBSERVATIONS OF GAMMA-RAY 
LINES FROM NOVAE.  I. LIMITS ON
THE POSITRON ANNIHILATION LINE IN FIVE INDIVIDUAL NOVAE}

\author{M. J. HARRIS\footnote{Universities Space Research Association,
harris@tgrs2.gsfc.nasa.gov}, J. E. NAYA\altaffilmark{1}, 
B. J. TEEGARDEN, T. L. CLINE, N. GEHRELS, 
D. M. PALMER\altaffilmark{1}, R. RAMATY, AND H. SEIFERT\altaffilmark{1}}
\affil{Code 661, NASA/Goddard Space Flight Center, Greenbelt, MD 20771}

\begin{abstract}

The Transient Gamma Ray Spectrometer (TGRS) on board the {\em WIND\/}
spacecraft has spent most of the interval 1995--1997 in a high-altitude
orbit where $\gamma$-ray backgrounds are low.
Its high-resolution Ge spectrometer is thus able to detect weak lines
which are slightly offset from stronger background features.  One such
line is predicted from nucleosynthesis in classical novae, 
where $\beta$-decays on a time-scale of a few hours in an
expanding envelope produce positrons that annihilate to generate a line 
which is blueshifted by a few keV away from the background annihilation
line at 511 keV.  The broad TGRS field of view contained five known
Galactic novae during 1995 January -- 1997 June, and we have searched
the spectra taken around the times of these events for the blueshifted
nova annihilation line.  Although no definite detections were made,
the method is shown to be sensitive enough to detect novae occurring on
ONeMg-rich white dwarfs out to about 2.5 kpc.

\end{abstract}

\clearpage

\section{Introduction}

In the standard thermonuclear runaway model of classical nova outbursts
the energy source is the explosive burning of H in a degenerate layer on
the surface of a white dwarf, composed of H-rich material accreted from
a companion contaminated by material taken up from the white dwarf by
diffusion.  The nuclear burning time-scale is rapid compared to the
lifetimes of certain key $\beta$-unstable isotopes in the reaction
chain, and substantial abundances of these nuclei are therefore built up.
The isotopes involved may be $^{7}$Be, $^{13}$N, $^{14,15}$O, $^{17,18}$F,
$^{22}$Na or $^{26}$Al, depending on the temperature and the chemical 
composition of the contaminating material from the white dwarf (CO-rich 
or ONeMg-rich).  Convection in the accreted layer carries the 
longer-lived ($>100$ s) of these 
species to the surface before they decay.  The $\gamma$-ray lines
resulting from decays at the surface are in principle observable, and would
provide a rather direct test of the model.

The $\beta$-unstable nuclei involved are all proton-rich and therefore
decay predominantly by emission of positrons.  Thus in addition to 
$\gamma$-ray lines from de-excitation of the daughter nuclei, we also
expect a line at 511 keV from the annihilation of the positrons with the
ambient material (Clayton \& Hoyle 1974,
Leising \& Clayton 1987).  Several previous experiments have searched for the
decay lines from the relatively long-lived isotopes $^{22}$Na and
$^{7}$Be, either from individual nearby novae (Leising et al. 1988, Iyudin
et al. 1995) or from the accumulated production of many novae in a 
wide field of view (Leising et al. 1988, Harris, Leising \& Share 1991,
Harris et al. 1997).  Here we report first
results of the first search for the 511 keV line, which reflects primarily
the production of shorter-lived isotopes such as $^{13}$N and $^{18}$F.

The Transient Gamma Ray Spectrometer experiment (TGRS) is very well
suited to the search for the 511 keV line, for several reasons.  First,
it is located on board a spacecraft whose orbit is so elliptical
that it has spent virtually all of its mission (1994 November--present)
in interplanetary space (the {\em WIND\/} mission).  In this
environment the $\gamma$-ray background level is relatively low, and 
interruptions due to passages of Earth's trapped radiation belts are
minimal (typically lasting a few hours at intervals of several months).
This is important because the bursts of 511 keV radiation are predicted
to occur only during a $\sim 6$ hr period at an uncertain interval before the
nova event (Leising \& Clayton 1987, Hernanz et al. 1997, and \S 2).
Second, the TGRS instrument is attached to the south-facing surface
of the rotating cylindrical {\em WIND\/} body, which points permanently
toward the south ecliptic pole.  The detector is unshielded, and TGRS
therefore has an unobstructed view of the entire southern (ecliptic)
hemisphere.  Third, and most importantly, the TGRS Ge detector has
sufficient spectral resolution to detect a 511 keV line which is slightly
Doppler-shifted away from the background 511 keV line which is always
present in space experiments, which arises from annihilation following 
$\beta^{+}$ decays of cosmic-ray spallation product nuclei.  This
background line is always very close to the rest energy,
whereas the most recent nova models predict the source line to be 
blueshifted by 2--5 keV (Hernanz 1997, private communication); by
comparison, the TGRS energy resolution at 511 keV has varied from
3--4 keV FWHM during the mission (Harris et al. 1998).
No previous experiment has possessed the same advantages of long and
continuous temporal coverage, broad spatial coverage, and fine
spectral resolution.

Our analysis procedure (described in \S 3) relies heavily on the
theoretically predicted properties of the 511 keV line.  There has not
been much incentive for theoretical work upon this line because of the
difficulty of resolving the nova line from the background line in the
previous generation of low-resolution experiments.
The early work of Clayton \& Hoyle
(1974) and Leising \& Clayton (1987) assumed white
dwarf models which were CO-rich, and treated the explosion parametrically.
Much fuller hydrodynamic models which can be applied at all stages from
accretion through explosion and nucleosynthesis have recently been
developed (Starrfield et al. 1992, Hernanz et al. 1996, Jose et al. 1997) 
and applied to the
positron annihilation problem (Hernanz et al. 1997, G\'{o}mez-Gomar
et al. 1998).  In this series of
models the largest 511 keV line fluxes come from the more massive
ONeMg-rich white dwarfs; indeed, fluxes as high as 
$10^{-2}$ photon cm$^{-2}$ s$^{-2}$ are expected in favorable cases.
For the purpose of analysis we shall in general assume the line properties of 
the Hernanz et al. (1997) ONe2 model, which generates this level of flux
over a $\sim 6$ hr period (\S 3).\footnote{ This model is referred to as A2
in G\'{o}mez-Gomar et al. (1998).}  In \S 2 we describe the difficulties which
are involved in a search for the 511 keV line around the times of those
novae which are known to have been in TGRS's field of view during the
mission.  Our results for the individual novae are presented in \S 4,
and we will show in \S 5 that for the ONe2 and other nova models TGRS 
sensitivity is good enough to justify a large-scale search for the line in
the full 1995--1997 TGRS data.

\section{Sample of Novae}

There were five known classical novae in the southern hemisphere between
1995 January and 1997 June, whose relevant properties are given in Table
1.  It should be noted that in many cases the observational material on
which they are based is extremely incomplete and their properties are often very
poorly known.  Columns 1 and 2 of Table 1 identify the nova.  Columns 3
and 4 indicate the day of discovery and the apparent magnitude 
$m_{V}(t=0)$ at discovery.  It is important to note that many novae are
probably discovered some time after visual maximum $m_{V}$, which
occurs $\sim$1--4 days after the explosion and 
the 511 keV line emission.  To bracket
the time during which the explosion must have occurred, we give the date
of the last prediscovery observation (or upper limit) in Column 5.

Observations during the early decline have been used to estimate the
speed class $t_{2}$, defined as the time necessary for $m_{V}$ to fall by
two magnitudes from the discovery value $m_{V}(t=0)$ (Column 6, Table 1).  
The typical situation is illustrated in Fig. 1.  Two facts are apparent from
the Table and the figure.  First, the sample of southern novae is likely
to be very incomplete.  Gaps of $> 20$ d in the coverage of the known
novae are common; therefore, very fast novae with $t_{2}$ of the order
of a few days could rise and then fall below the detectable level during
such a gap.  Warner (1989) estimates that this incompleteness could
affect even novae so bright as $m_{V} \sim 3$.  Second, the distance
to any nova can only be determined by comparing the apparent magnitude
at some known epoch (usually at visual peak) with an absolute
magnitude $M_{V}$.  Empirical formulae are available which relate the absolute
magnitude $M_{V}$ to the speed class $t_{2}$ (Warner 1995).
However, the peak $m_{V}$ is unknown since the
time of the peak is unknown, and it can only be estimated by
extrapolating back up the prediscovery light curve (dashed line in Fig. 
1).  Thus the estimated distance $r$ to the nova is a function 
$r(t)$ of the time before discovery at which the outburst is assumed to occur.

The relationships involved are the well-known distance modulus formula
including extinction $A_{V}$
\begin{equation}
m_{V} - M_{V} = 5 \log r - 5 + A_{V}        
\end{equation}
and the extrapolation up the prediscovery light curve
\begin{equation}
m_{V}(t) = m_{V}(t=0) + 2 \frac{t}{t_{2}}     
\end{equation}
which can be combined to yield a transcendental relation for the 
estimated distance $r(t)$ 
\begin{equation}
\log r(t) + \frac{A_{V}(r)}{5} = 1 + \frac{m_{V}(t) - M_{V}}{5}  
\end{equation}
if $M_{V}$ and $A_{V}(r)$ are given.  We solved these equations numerically
for each nova during each prediscovery interval using an empirical 
relation between $M_{V}$ and $t_{2}$
\begin{eqnarray}
M_{V} & = & 2.41 \log t_{2} -10.7 ~~~\mbox{for 5 d $< t_{2} < 50$ d} \\
 & = & -9 ~~~\mbox{for $t_{2} \le 5$ d} \\ 
 & = & -6.6 ~~~ \mbox {for $t_{2} > 50$ d}
\end{eqnarray}
(Warner 1995) and the prescription for extinction as a function of celestial 
coordinates and of distance given by Kil'pio \& Malkov (1997).  The range 
of distances given in Column 7 of Table 1 for each nova was thus obtained.  
In all cases the smallest value holds true if the outburst occurred just after
the last prediscovery measurement (Column 5) and the largest value
corresponds to an outburst occurring at the time of discovery (Column 3).
We neglect uncertainties due to the scatter in the empirical relation 
Eqs. (4--6), which are substantial, but still are less important than
the uncertainty in the outburst epoch for relatively fast novae such
as most of those in Table 1.

Very little information appears to be available on the abundances of
the novae in Table 1.  Woodward et al. (1998) have shown that N Cru 1996 
was definitely of the neon subclass, involving an ONeMg white dwarf.  More 
tentatively, enhancement of N but not of O in N Cir 1995 
(Greeley, Blair \& Long 1995) suggests that it might have occurred on a
very massive ONeMg white dwarf (Politano et al. 1995).  Since theoretical models
(Hernanz et al. 1997, G\'{o}mez-Gomar et al. 1998) suggest that the 511 keV 
line is much more likely to be detectable from neon novae, both
these novae might be promising candidates for
detection if they are at the nearer ends of the ranges in Table 1 Column 7.
 
\section{Observations and Analysis}

\subsection{Observations}

Since launch in 1994 November the {\em WIND\/} spacecraft has been
raised into a complicated and extremely elliptical orbit,
most of which is spent outside Earth's magnetosphere.
Perigee passes taking the spacecraft within the magnetosphere (and 
therefore exposing it to high charged particle fluxes in the trapped 
radiation belts) are brief (typically $\leq 1$ d).  The overall
level of background $\gamma$-radiation is therefore
relatively low, being mainly due to the cosmic diffuse background and
the effects of irradiation by the Galactic cosmic rays.

The TGRS detector is a radiatively cooled 35 cm$^{2}$ Ge crystal
sensitive to energies between 20 keV--8 MeV.  It is mounted on the top surface
of the {\em WIND\/} spacecraft body, which permanently faces 
the south ecliptic pole; the spacecraft is stabilized 
by rotation in the ecliptic plane with a 3 s period (Owens et al.
1995).  Although the
detector is unshielded, some capability for localizing sources exists 
along the ecliptic plane, since a passive Pb occulter is fixed 
concentric with the detector and carried around it by the spacecraft's
rotation, thereby modulating signals from the rotation plane with a 3 s
period (see figures in Teegarden et al. 1996).

The database of regular TGRS background spectra consists of count
rates between 20 keV and 8 MeV which have been accumulated continuously 
in 1 keV energy bins during 24 min intervals since launch, except for 
the epochs of perigee passages.  The data stream is also interrupted by
memory readouts of triggered data from solar flares and 
$\gamma$-ray bursts for periods of $\sim 2$ hr (Owens et al.
1995).  We summed these
spectra into 6 hr intervals, corresponding to the expected time-scale
of the signal from a classical nova (Hernanz et al. 1997).  As noted
in \S 2 above, it is frequently unclear how much time elapsed between
the explosion and the visual discovery for any given nova.  We were
therefore usually obliged to take 6-hr spectra for the entire period
prior to discovery, as far back as the latest available prediscovery
observation (dashed line, Fig. 1).  Of course, the implied
apparent magnitude of the nova brightens as one proceeds back along
the light curve, and in some cases this in turn implies that a very
bright nova must somehow have escaped detection.  We therefore did not
consider epochs prior to discovery for which $m_{V}$, extrapolated
according to speed class and corrected for extinction, would have fallen
below 3.0 (cf. Warner 1989).

Two of the novae listed in Table 1 lay in the approximate direction of
the Galactic center, near where the Galactic plane crosses the ecliptic,
and so were occulted by the Pb occulter
(N Sgr 1996 and N Sco 1997).  The TGRS occulted data are returned
in four broad 64-channel energy windows, only one of which is binned at
1-keV resolution; this window covers the region 479--543 keV and 
therefore contains the blueshifted nova line predicted by Leising \&
Clayton (1987) and Hernanz et al. (1997).  For 
N Sgr 1996 and N Sco 1997 we therefore summed the occulted 
spectra over the same prediscovery time intervals as described above
for background spectra.  We thus expected to have an independent check
(by means of location) on any positive signal discovered in the background
spectra.

\subsection{Analysis of Background Spectra}

A characteristic TGRS background spectrum in the range 490--530 keV
during a 6-hr interval is shown in Fig. 2.  The overall features of such
spectra, which are very stable from one spectrum to the next,
are the following: first, a strong line due to positron annihilation in
the spacecraft and the passive material surrounding the
detector.  This line is expected to lie at the exact
rest energy 511.00 keV\footnote{~The instrument gain
has varied slightly (by $\sim 0.5$\%) during the mission.  For the
purpose of measuring line positions close to 511 keV, we can use the
position of this background annihilation line as a reference energy to
correct this gain shift.  The procedure involved --- fitting
a Gaussian line and equating the position of its peak to 511.00 keV --- is
described by Harris et al. (1998).}.  The line shows a clear asymmetry
in the form of a broad red wing of instrumental origin. This asymmetry is 
probably caused by charge collection losses; the increasing effects of 
radiation damage have caused it to become more pronounced as the
mission has proceeded.  Second, there is an underlying continuum at 
energies above and below the background line energy,
which over a short range of spectrum may be approximated by a power
law.  Third, there is a discontinuity in the continuum at the line energy
which is due to incomplete absorption of the line photons in the detector.

The principle behind our analysis was to fit the selected background spectra
(see \S 3.1) with such components as those described above, plus lines
fixed at the energies and widths predicted by theory from neon novae.
Significant fitted amplitudes in these lines would be taken as detections.  
The predicted lines are described in Table 2; they correspond to the same
annihilation line at two different epochs, 6 and 12 hours after the
explosion [Hernanz et al. 1997, Hernanz 1997 (private communication),
G\'{o}mez-Gomar et al. 1998].
We will refer to these as the "6-hr" and "12-hr" lines respectively.  Note
that, while the 6-hr line is quite well separated from the background
line, the 12-hr line is only blueshifted from it by 2 keV and will thus be 
badly blended, even with TGRS's superior energy resolution\footnote{As
measured by Harris et al. (1998) the FWHM 
energy resolution of TGRS at energy 511 keV degraded during the mission
from 3.2 keV to $> 4$ keV due to radiation damage from cosmic-ray 
impacts; this loss of resolution appears to be due to the 
growth of red wings on spectral lines as described below.}.  We thus
expect worse statistical and also
systematic errors on the 12-hr line.  The detection and measurement of the nova
line signal (if any) thus depend almost entirely on
our results for the 6-hr line.

Fitting the full spectrum shown in Fig. 2 with the full
complement of model components (background 511 keV Gaussian line; 
nova 6-hr or 12-hr Gaussian line\footnote{ The fits are not strongly
sensitive to the precise line shape, so long as the FWHM is fixed at 8 keV
(Table 2).  For example, Leising and Clayton (1987) argued in favor of
the broad flat-topped line shape expected from the near side of an 
expanding shell.  We tested this line shape and found little difference
from results obtained with the Gaussian shape which Hernanz (1997, private 
communication) used as an approximation.}; power law; step 
function at 511 keV) does not lead to acceptable values of 
$\chi^{2}$ (e.g. $\simeq 500$ for 35 degrees of freedom in Fig. 2).  
The main reason for this is the asymmetric
shape on the red side of the background 511 keV line which is apparent
in Fig. 2, whose origin is unclear.  It became more pronounced as the
mission proceeded, and was therefore probably due to accumulated radiation
damage; an investigation of this effect is in progress (Kurczynski et al.
1999). 

Although the anomalous red component of the background 511 keV line is
not perfectly understood, this is not of crucial importance to our
search for excess emission on the {\em blue\/} side of the profile.
We have found that the best fits are obtained when a simpler model
(background 511 keV line; 
nova 6-hr or 12-hr line; constant approximation to power law) is
fitted to energies $\geq 511.0$ keV only (Fig. 3a, b).  The
values of $\chi^{2}$ for such fits typically lie between 15--25 for 17
degrees of freedom.  

In detail, our method of fitting each spectrum was to vary
the model parameters until the count rate was best reproduced
according to the minimum of $\chi^{2}$; errors were calculated
by the method of Lampton, Margon \& Bowyer (1976) for mapping the
range of parameter space where the minimum $\chi^{2}$ value is
exceeded by 1.  The line amplitudes from the count rate were normalized by
the detector effective area (a function of the direction to the nova), 
which is known from Monte Carlo simulations (Seifert et al. 1997).  In
Fig. 3b (dotted line) we use this detector response to show that a neon
nova with the theoretically expected properties would be easily detectable
at a distance 1 kpc.

A typical series of fitted amplitudes for the 6-hr line, during times
when N Sco 1997 may have erupted, is shown in Fig. 4a.  Three features
of this result call for attention.  First, the points are 
distributed about a mean value in a way consistent with random scatter
($\chi^{2}$ per degree of freedom = 0.71).  
Second, (consistent with this) there is no individual
point of high significance relative to the mean such as would be expected 
from a real nova line.  Third, the mean value is not zero --- there is
a systematic positive nova line amplitude.

The systematic positive amplitude is probably due to a systematic departure
of the background 511 keV line from Gaussian shape on its blue wing (for
example, if the anomalous red component seen in Fig. 2 has a Gaussian
shape, its blue tail would effectively contribute such 
a constant value to the nova line).  We have found in all cases that the
systematic positive offset is very stable.  It has increased slowly
during the mission; the time-scale of the increase ($\sim$ months to 
years) is much longer than the time scales ($\sim 10$ days) over which we 
search for lines from a given nova.  In this respect the behavior of 
the positive offset is very similar to that of other measures of
degradation of detector performance, such as energy resolution
(Harris et al. 1998) and the amplitude of the anomalous red component
seen in Fig. 2.

Apart from this stable and well-behaved offset (which can be removed
from the nova line flux by simple subtraction) there appear to be few
or no systematics in our nova line measurements made from background
spectra.  We therefore believe that single significant positive 
measurements (following subtraction of the systematic offset)
would be valid evidence for short-lived ($\sim 6$ hr) line emission.
We now turn to two internal checks which could be made on such a detection.
In \S 3.3 we examine the usefulness of data modulated by the TGRS
occulter for some celestial directions.  In \S 3.4 we show that small
improvements in sensitivity can be obtained from measuring the 511 keV line
at epochs subsequent to its peak flux at $t = 6$ hr (what we have called
the "12-hr line").

\subsection{Analysis of Occulted Spectra}

The Pb occulter referred to in \S 3.1 subtends an angle of
$90^{\circ}$ along the
ecliptic and approximately $16^{\circ}$ across it, and is 1 cm in
thickness.  The spectra in 64 1-keV energy channels around 511 keV,
when modulated with a 3 s period by the occulter, are binned into 128 angular
channels ("sectors") of $2.8125^{\circ}$ in ecliptic longitude.
We obtained spectra for N Sgr 1996 and N Sco 1997 by fitting occultation
dips centered on the respective ecliptic longitudes 
($\lambda = 275.66^{\circ}$ and 
$268.73^{\circ}$) to each energy channel (cf. Teegarden
et al. 1996, Harris et al. 1998).  A typical spectrum for a 6 hr
period is shown in Fig. 5.

All background features varying on time-scales $> 3$ s are subtracted
out of such a spectrum, notably the 511 keV background line which is so
prominent in Fig. 2.  The subtraction is very effective and the residual
spectrum in Fig. 5 is almost featureless.  For occultations having phase
close to that of the Galactic center ($\lambda = 266.8^{\circ}$), the
strongest feature in the residual spectrum is expected to be the
diffuse Galactic 511 keV annihilation line (see e.g. Purcell et al.
1997).  However the measurement by Harris et al. (1998) with the
TGRS occulter during 90 d intervals indicates that during 6 hr intervals
this line will be detected at a level of only $\sim 0.35 \sigma$.

We therefore fitted the occulted spectra
between 490--530 keV by a model containing two
lines, one at exactly 511 keV due to the diffuse Galactic source, and one 
at either the 6-hr or 12-hr line energy (Table 2).  The gain-corrected
511 keV line positions were taken from the corresponding background
spectra (see footnote, \S 3.2).  The 511 keV line widths were fixed
at the values measured by Harris et al. (1998) during 90 d intervals
containing the nova epoch.

Results from a single 6 hr period for the 6-hr line in
N Sco 1997 are shown in Fig. 5 (solid, dashed and dot-dashed lines); in
Fig. 6 we show the time series of all such fits for this nova.  
Although there do not appear to be any large systematic effects, it is
immediately apparent that some of the 6 hr intervals exhibit anomalously
large errors.  These situations occur in a model which possesses 
multiple minima
of $\chi^{2}$ lying close to each other in the model's parameter space.
The region defined by $\chi^{2} +1$ is then much larger than would naively
be expected from the behavior of $\chi^{2}$ in the region around the true
minimum.  In the present case, which is a
two-line model where the lines are not well resolved, such situations
tend to arise when the spectrum
is best fitted by lines with amplitudes 
of opposite signs.  We found this to be true of the badly-behaved 
fits in Fig. 6.

\subsection{Combination of 6-hr Line and 12-hr Line Results}

The time series of background spectrum fit results for N Sco 1997 are
shown in Fig. 4a for the 6-hr line and in Fig. 4b for the 12-hr line.
We see that the statistical errors on the 12-hr line are generally
much larger, as expected, due to worse blending with the
511 keV background line.  The systematic positive offset described
in \S 3.2 for the 6-hr line is also worse in the
12-hr line flux measurement.  

The 6-hr and 12-hr line fits are independent and can be combined so
as to improve the significance of a detection in one of them.  In our
simple approximation, they are combined with a weight proportional to
their contribution to the Hernanz et al. (1997) light-curve (c.f. Table
2) and proportional to the inverse square of the statistical
error, in the usual way.  Each 6-hr line measurement (Fig. 4a) is
combined with the 12-hr line measurement from the following 6-hr
interval (Fig. 4b).  The results for N Sco 1997 are shown in
Fig. 4c for background spectrum fits\footnote{In this figure, and from
here onward, the positive offset in these measurements is subtracted
off.}

It is clear from Fig. 4b that the quality of the 12-hr line
measurements is much worse than that of the 6-hr lines, as expected.
Thus the 12-hr line measurements yield only
a small gain in the sensitivity of the line search, when 
combined with the 6-hr line measurements which contain most of the
information.  This may be illustrated by the similarity between Figs. 4a 
and 4c; clearly very little information was added by
including the 12-hr line.

\section{Results}

Our results for N Cir 1995, N Cen 1995, N Sgr 1996, and N Cru 1996 
are shown in Figs. 7--10.  For N Sgr 1996 there are both background and
occulted spectra (Figs. 9a and 9b), for the others only background
spectra.  The results are given as the nova line flux at all epochs
when the thermonuclear outburst could have occurred; note that (as
derived in \S 2) the theoretical value with which they are compared
is time-dependent (dotted line).
The positive systematic offsets described in \S 3.2 have been
subtracted.  In the absence of any single strong positive 6-hr line
detections, we have combined the 6-hr and 12-hr line results as 
described in \S 3.4.

Once the positive offset has been removed, the distribution of the 
measured line intensities is quite free from systematics,
as previously deduced from the N Sco 1997 results (\S 3.2 and Fig. 4c).
The distribution of the complete ensemble of measurements (264 points)
agrees very well with a random distribution about zero.
We therefore conclude that none of the novae in our sample emitted the
expected blueshifted 511 keV line, down to a limiting flux which varied
for each nova.  We express this limit
conservatively as the $3 \sigma$ upper limit from the statistical
errors on each 6-hr point; it is represented by 
dashed lines in Figs. 4c and 7--10, and the mean values for each nova
are given in Table 3.  The mean values vary slightly from nova to nova
partly because of the slow degradation of detector performance (\S 3.2), 
but mainly due to the variation of effective area with angle of incidence
of the nova in the detector.

The typical $3 \sigma$ sensitivities in Table 3 may be compared with
the prediction of $1.6 \times 10^{-2}$ photon cm$^{-2}$ s$^{-1}$ for
a neon nova at 1 kpc (Table 2).  From this we derive our most important
conclusion, namely that {\em the TGRS measurements are sensitive enough
to detect neon novae to a distance of about 2.5 kpc at $3 \sigma$
significance\/}, for favorable
estimates of the nova line emission (Hernanz et al. 1997, model
ONe2).

There remains a small possibility that a 511 keV line event associated
with one of the novae has been
missed due to incomplete coverage.  It can be seen from Figs. 4 and
6--10 that losses of entire 6-hr intervals are rare, generally involving
perigee passes of $\le 1$ d (Figs. 7, 8, 10).  Briefer losses of data 
within a few 6-hr intervals cause larger than normal error bars, and
arise when TGRS's operation mode changes following a $\gamma$-ray
burst trigger, or from telemetry errors.  Overall, 
we estimate that the instrument's temporal
coverage at the characteristic sensitivities in Table 3 (2--3
$\times 10^{-3}$ photon cm$^{-2}$
s$^{-1}$ in 6 hr at $3 \sigma$) is 80--85\%, and that
$\sim 97$\% of all 6 hr intervals contain at least some live time.

\section{Discussion}

\subsection{Results for Individual Novae}

As emphasized in \S 2, it is very difficult to use our results for
the individual novae to constrain the ONe2 model because of our
lack of knowledge of two related nova parameters, the epoch of
explosion and the distance.  In Figs. 4c and 7--10 we use the
limited information given in Table 2 to plot the
predicted fluxes $\phi_{pred}$ as dotted lines, which depend on
the time relative to discovery according to
\begin{equation}
\phi_{pred}(t) = 0.016/r(t)^2
\end{equation}
where the distance $r(t)$ in kpc is obtained from Equation (3).
Note that $\phi_{pred}$ is {\em not\/} a light curve;
as shown in Table 1, there is usually a wide range of
allowed explosion times, and $\phi_{pred}$ is the predicted flux
{\em at any one time\/} of explosion.  In two cases
(N Sco 1997 and N Sgr 1996; Figs. 4c and 9a) the predicted flux
is always below our $3 \sigma$ upper limit, for any of the allowed
explosion times; these novae must be too distant for our results 
to be used to constrain even the most optimistic nova model (Hernanz et al.
1997, model ONe2).  The other three novae (N Cir 1995, 
N Cen 1995 and N Cru 1996)
are predicted to yield a detectable $\phi_{pred}$ only if
the explosion occurred early in the allowed range of times [prior to
$\sim 12$ d before discovery for N Cir (Fig. 7), 7 d before discovery 
for N Cen (Fig. 8), and 9 d before
discovery for N Cru (Fig. 10)].  In other words, the allowed
range of distances for these novae includes the TGRS detection
limit $\sim 2.5$ kpc.  The ONe2 model can be tested using
our results for these novae if further investigations yield either
the appropriate constraints on explosion time, or an independent
measurement of the distance to the nova.  

\subsection{Implications for Nova Searches}

The data analyzed in this paper covers about 70 d, representing only
a small fraction of the whole TGRS database.  An obvious extension of
this work would be to analyze the entire database in order to
detect, or place upper limits upon, the occurrence of all neon
novae lying within the 2.5 kpc radius estimated in \S 4.  There
are good reasons for believing that many novae within this radius
escape detection.  As noted in \S 2, visual searches for novae
are usually episodic, containing gaps in temporal coverage within
which fast-rising and decaying novae are likely to be missed;
coverage is particularly uneven in the southern hemisphere.  
Further, the southern Galactic plane contains regions of very strong
interstellar absorption; $A_{V}$ values up to 4 are found within 2.5
kpc in some directions (Neckel \& Klare 1980).  There is therefore
considerable uncertainty, not only about the global Galactic nova
rate, but also about its spatial distribution; novae in a disk
population may be more common than the spheroidal population modeled
by Leising \& Clayton (1985), but less observable due to obscuration
(Hatano et al. 1997).  There is evidence that the neon subclass may be
more concentrated toward the plane, and thus more under-represented
in global nova statistics (Della Valle et al. 1992).

Estimates of the success rate of a $\gamma$-ray line search must
be very uncertain, for the above reasons; in fact, the best
justification for such a search is precisely that it is the only
way to obtain an unbiased census of novae on which to base 
global nova rates.  Estimated rates for the Galaxy as a whole range
from 11--260 yr$^{-1}$, tending to cluster about values
35--50 (Shafter 1997), with about one-third occurring on
ONeMg white dwarfs (Gehrz et al. 1998).  To estimate TGRS's rate of
detections, we note that a full-scale search of almost 3 years of TGRS
data will yield a much higher chance detection rate than that found
in the present work ($< 1$ detection in 70 d), so that the threshold
for detection must be set higher than the value $3 \sigma$ used here.
Since the distribution of our null measurements is very close
to the normal distribution expected by chance, it is easy
to show that a threshold level $\simeq 4.6 \sigma$ yields a probability
$< 1$\% of a single false detection by chance (Abramowitz \& Stegun 1964).
This threshold level corresponds to a distance $\simeq 2$ kpc for a
neon nova.  A simple model of nova distribution following star
counts (Bahcall 1986, integrating the disk over $z$)
suggests that $\sim 2$\% of Galactic novae occur within 2 kpc of the
Sun in the southern hemisphere.  Moderate estimates of the global 
nova rate $\sim 50$ yr$^{-1}$ then imply that TGRS would detect
one neon nova during $\sim 3$ yr; higher global rates, or a higher
fraction of neon novae, would imply more detections. 
 
Further papers in this series will utilize this search strategy, and 
will follow up the possibilities arising from detailed study of the
already known novae as described in the previous section.

\acknowledgments

We acknowledge the work of the small but dedicated teams who monitor
southern hemisphere novae: the Royal New Zealand Astronomical Society
(communicated by F. Bateson) and the VSNET network of variable star
observers (communicated by T. Kato).  We are grateful to Dr. M. Hernanz for
providing pre-publication results and to Drs. C. Woodward and M.
Greenhouse for helpful discussions.
Theresa Sheets (LHEA) and Sandhia Bansal (HSTX) provided
assistance with the analysis software.

\clearpage

\begin{figure}

\caption{Schematic nova visible light curve.  Full line --- observed
light curve after discovery.  Dashed line --- unobserved light curve
before discovery; note that the position of peak $m_{V}$ is unobserved.
Last pre-discovery observation is labeled "P".  Time to fall by 2 
magnitudes is labeled $t_{2}$ ("speed class").} 

\caption{Characteristic TGRS background count spectrum at energies around
511 keV, obtained prior to the discovery of N Sco 1997 during the
interval 3 June 12h--18h UT.  Full line --- best-fitting model spectrum
incorporating power law, step function at 511 keV, and an asymmetric
511 keV background line fitted by two Gaussian components (broad on
the red wing, narrow on the blue wing).}

\caption{(a) Our preferred fit to the blue wing of the TGRS
background line at 511 keV; same data as in Fig. 2 with
the power law continuum subtracted.  Apart from the power law, the
components of the fit are a line with the 
width and position of the nova 6-hr line in Table 2
(dot-dashed line) and a Gaussian line fitting the blue wing of the
511 keV background line (dashed line).  The total model spectrum
is the full line.  (b)  Expansion of Fig. 3a showing the significance of the
fitted nova 6-hr line (dot-dashed line of amplitude $5.2 \pm 0.1 \times
10^{-3}$ photon cm$^{-2}$ s$^{-1}$; other symbols as in Fig. 3a).
Also shown is the theoretically expected level of the nova 6-hr line for
a nova at 1 kpc (Hernanz et al. 1997 model ONe2, dotted line).}

\caption{(a) Measured fluxes in the nova 6-hr line for N Sco 1997
before outburst, from fits to the spectral model illustrated in Fig. 
3b.  (b) Fluxes for N Sco 1997 in the 12-hr line measured by the 
same procedure.  (c) Combined fluxes for N Sco 1997 in the
6-hr and 12-hr nova lines, expressed as equivalent 6-hr line fluxes
assuming the light curve predicted by Hernanz et al. (1997).  The 
positive offsets in {\em (a)\/} and {\em (b)\/} have been subtracted.
Dashed line --- $3 \sigma$ upper limits on the line flux.  Dotted
line --- predicted flux from the nova parameters in Table 1 and the
ONe2 model of Hernanz et al. (1997), discussed in \S 5.1.}

\end{figure}

\begin{figure}

\caption{Characteristic TGRS count spectrum from the direction of N 
Sco 1997 as modulated by the occulter, during the same 6-hr period
as in Figs. 2 and 3.   Symbols for the model fit are 
the same as in Fig. 3b: dashed line --- Gaussian fit to 
the residual 511 keV background line.  Dot-dashed line --- Gaussian
with nova 6-hr line parameters.  Full line --- total model spectrum.
Dotted line --- theoretically expected level of the nova line at a
distance 1 kpc (Hernanz et al. 1997, model ONe2).}

\caption{Measured fluxes for N Sco 1997 in the nova 6-hr line from 
occulted spectra (as fitted in Fig. 5).}

\caption{Measured fluxes for N Cir 1995 in the 6-hr nova line
(combined with 12-hr measurement as described in \S 3.4),
from background spectra.  Symbols as in Fig. 4c: dashed 
line --- $3 \sigma$ upper limits on flux.  Dotted line --- predicted
flux from the nova parameters in Table 1 and the Hernanz et al. (1997)
ONe2 model.}

\caption{Measured line fluxes for N Cen 1995.  Symbols as in
Fig. 7.}

\caption{(a) Measured line fluxes for N Sgr 1996 from background
spectra.  Symbols as in Fig. 7.  (b) Measured line fluxes for N Sgr 1996
from occulted spectra.}
  
\caption{Measured line fluxes for N Cru 1995.  Symbols as in
Fig. 7.}

\end{figure}

\clearpage

\begin{table*}
\begin{center}
\begin{tabular}{llcccccl}
\tableline
Nova & Alias & Date of & $m_{V}$ at & Last pre- & $t_{2}$, & 
Estimated & References \\
 & & discovery & discovery & discovery & d & distance, pc \\
\tableline
N Cir 1995 & & Jan 27.328 & 7.2 & Jan 12.0 & $\sim 20$ & 2830--4790 & 
1,2,3,4 \\
N Cen 1995 & V888 Cen & Feb 23.31 & 7.2 & Jan 27 & $\sim 6$ & & 2,4,5 \\
 & & & & Feb 11.125\tablenotemark{a} & & 1390--8330 & \\
N Sgr 1996 & V4361 Sgr & Jul 11.527 & 10.0 & Jun 19.631 & 53 & 6660--9810 &
2,6 \\
N Cru 1996 & CP Cru & Aug 26.98 & 9.25 & Aug 7.0 & 5.2 & & \\
 & & & & Aug 12.125\tablenotemark{a} & & 930--18000 & 2,4,7 \\
N Sco 1997 & & Jun 5.09 & 8.5 & Jun 2.09 & 4.5 & 6830--17740 & 2,8 \\
\tableline

\end{tabular}
\end{center}

\tablenotetext{a}{Epoch of $m_{V} = 3$ extrapolated back from discovery.
See text, \S 3.1. \\
{\em REFERENCES\/} \\
1. Liller 1995a.\\
2. VSNET 1998.\\
3. Greeley, Blair, \& Long 1995.\\
4. Bateson 1998.\\
5. Liller 1995b.\\
6. Sakurai 1996.\\
7. Liller 1996.\\
8. Liller 1997.\\}

\caption{Novae observable by TGRS, 1995--1997}

\end{table*}

\clearpage

\begin{table*}
\begin{center}
\begin{tabular}{cccc}
\tableline
Line energy, & Line width & Time from & Theoretical intensity at \\
keV\tablenotemark{a} & FWHM, keV\tablenotemark{a} &
explosion, hr\tablenotemark{a} & 1 kpc, $\gamma$ cm$^{-2}$ 
s$^{-1}$~\tablenotemark{b} \\
\tableline
516 & 8 & 6 & $1.6 \times 10^{-2}$ \\
513 & 8 & 12 & $7.8 \times 10^{-3}$ \\
\tableline

\end{tabular}
\end{center}

\tablenotetext{a}{ Hernanz 1997, private communication.}
\tablenotetext{b}{ Hernanz et al. 1997, best-case assumption (model ONe2).}

\caption{Nova 511 keV line properties expected at different epochs}

\end{table*}

\clearpage

\begin{table*}
\begin{center}
\begin{tabular}{lcc}
\tableline
Nova & Angle of incidence, & Mean $3 \sigma$ upper limit in 6 hr, \\
 & degrees & photon cm$^{-2}$ s$^{-1}$ \\
\tableline
N Cir 1995 & 44.9 & $2.2 \times 10^{-3}$ \\
N Cen 1995 & 42.0 & $2.0 \times 10^{-3}$ \\
N Sgr 1996 & 95.2 & $2.8 \times 10^{-3}$ \\
N Cru 1996 & 36.9 & $2.3 \times 10^{-3}$ \\
N Sco 1997 & 83.4 & $2.9 \times 10^{-3}$ \\
\tableline

\end{tabular}
\end{center}

\caption{Upper limits on 511 keV line emission from novae}

\end{table*}

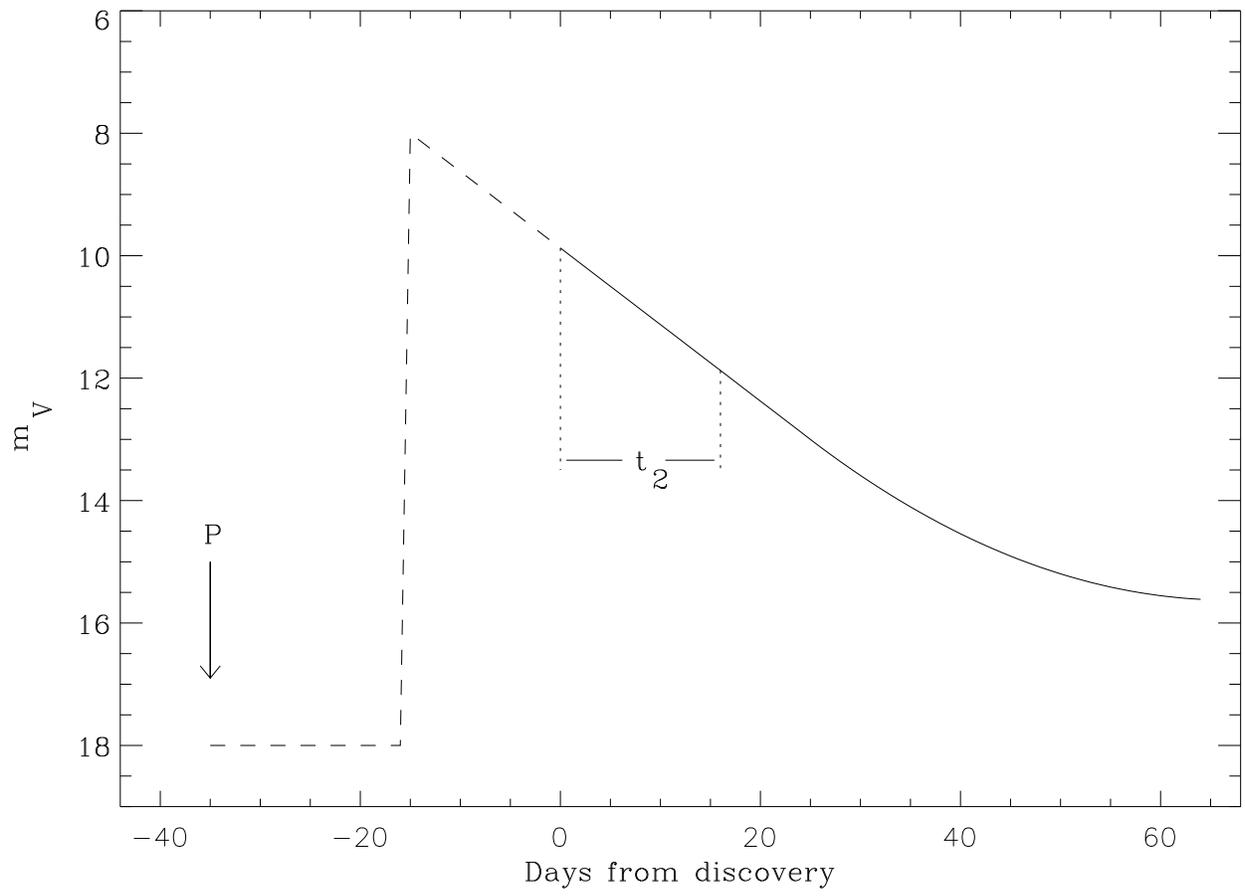
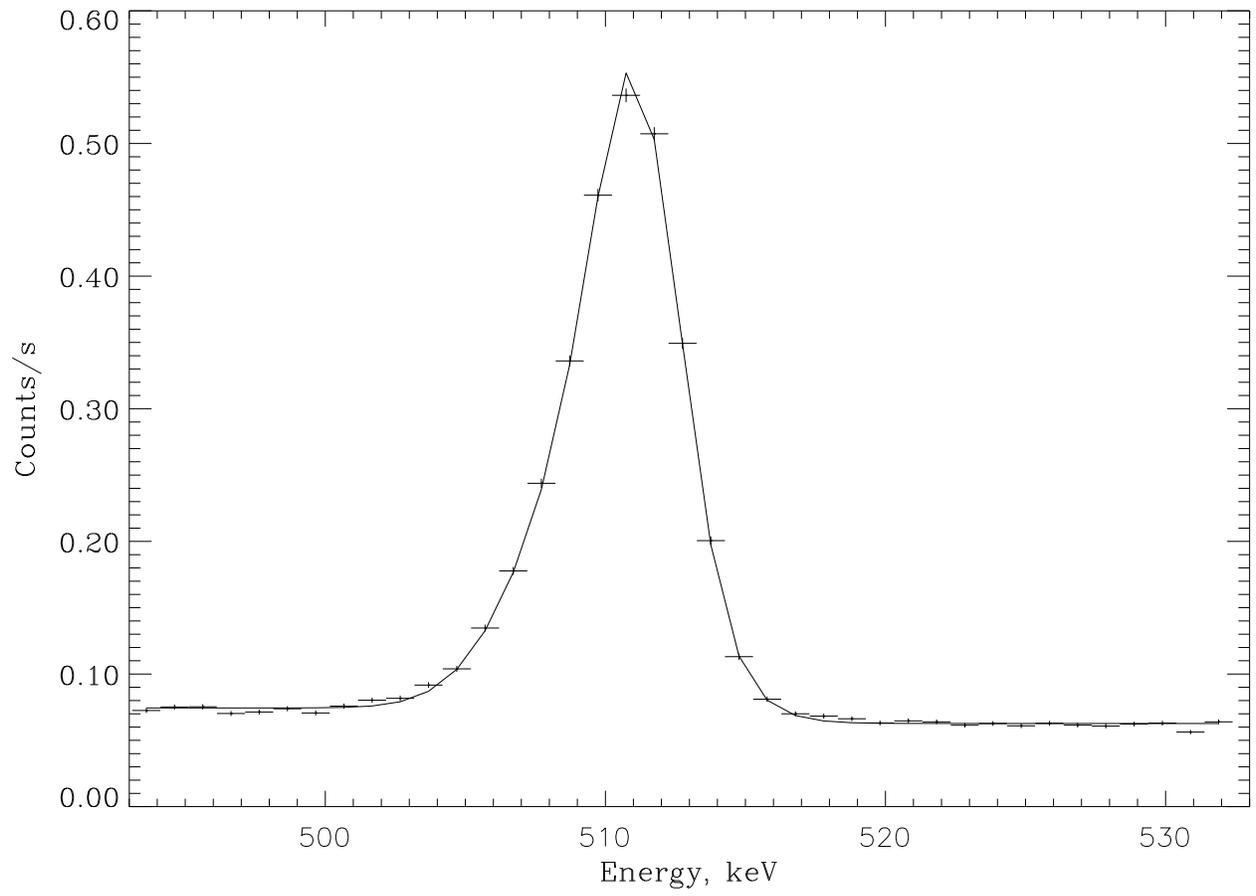
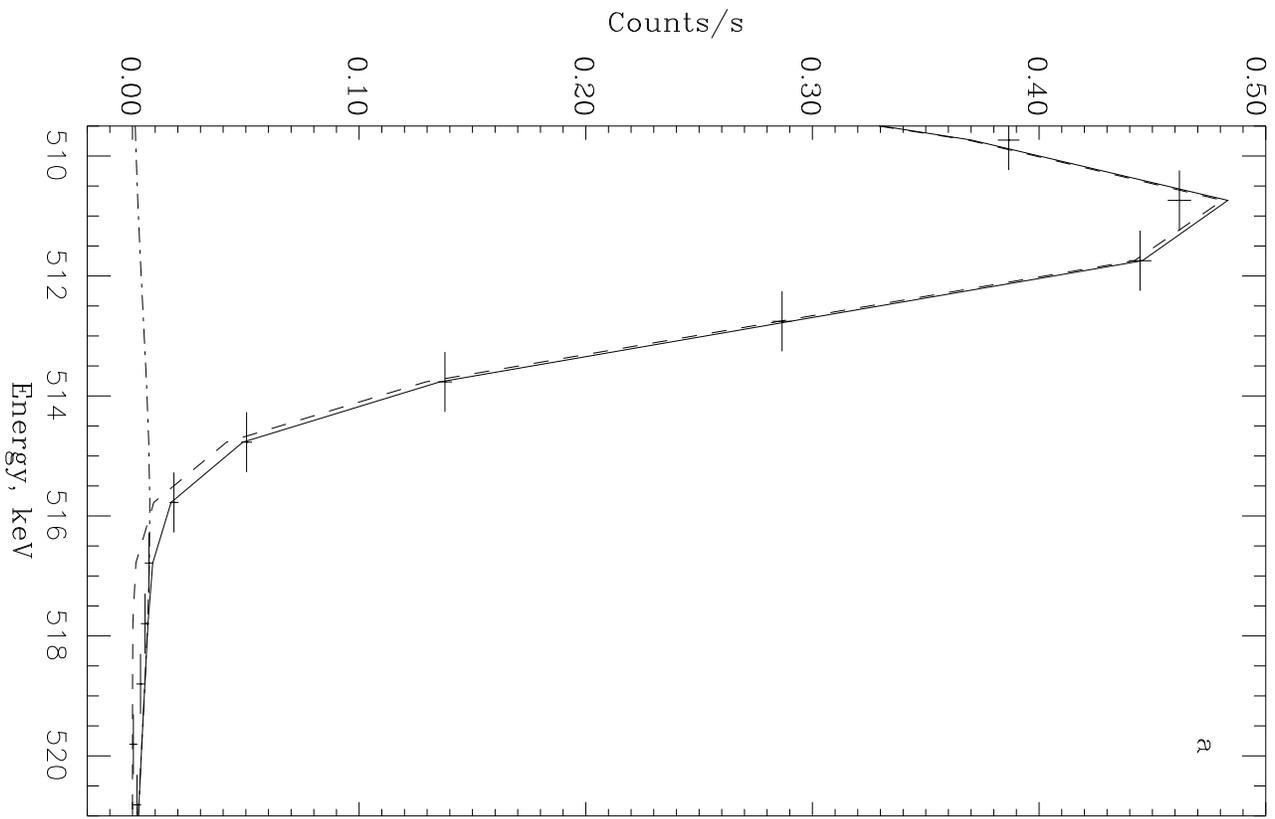
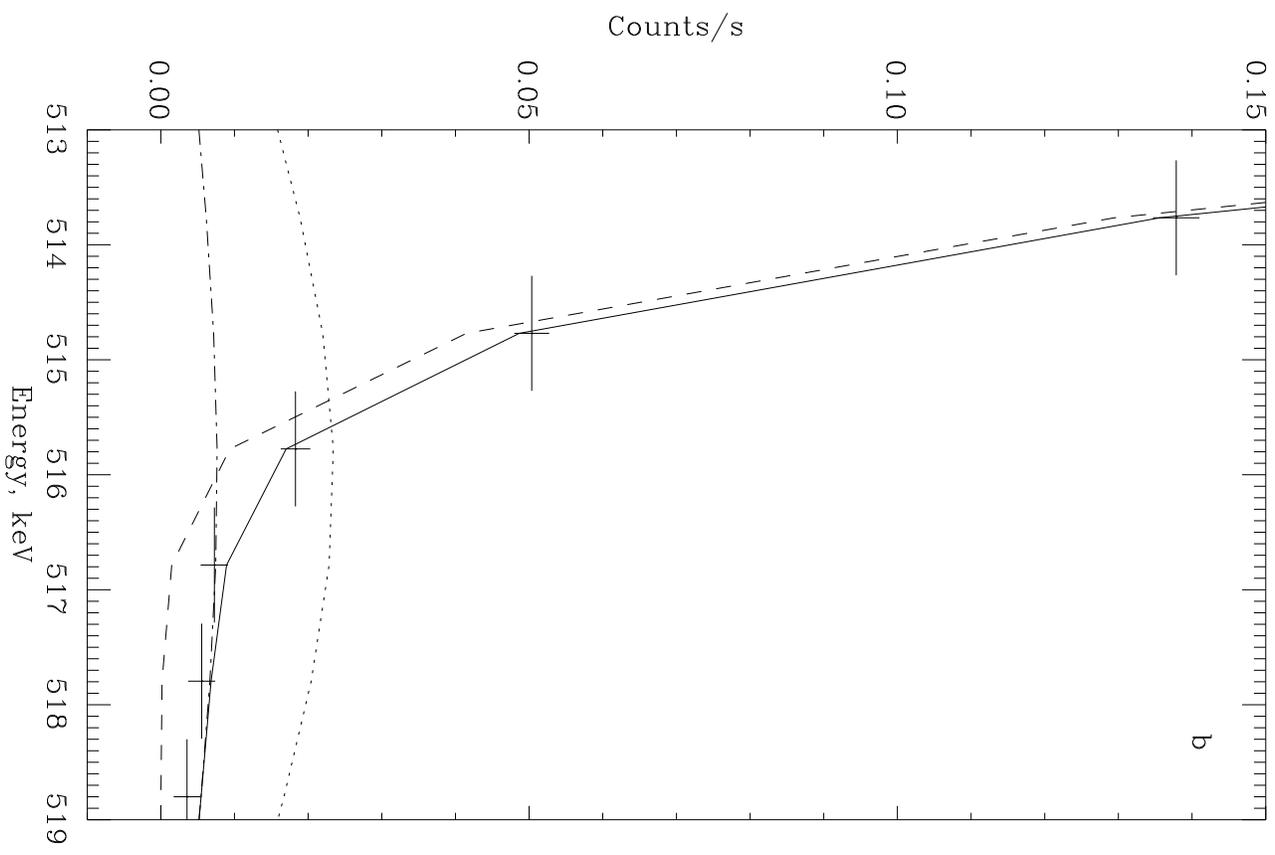
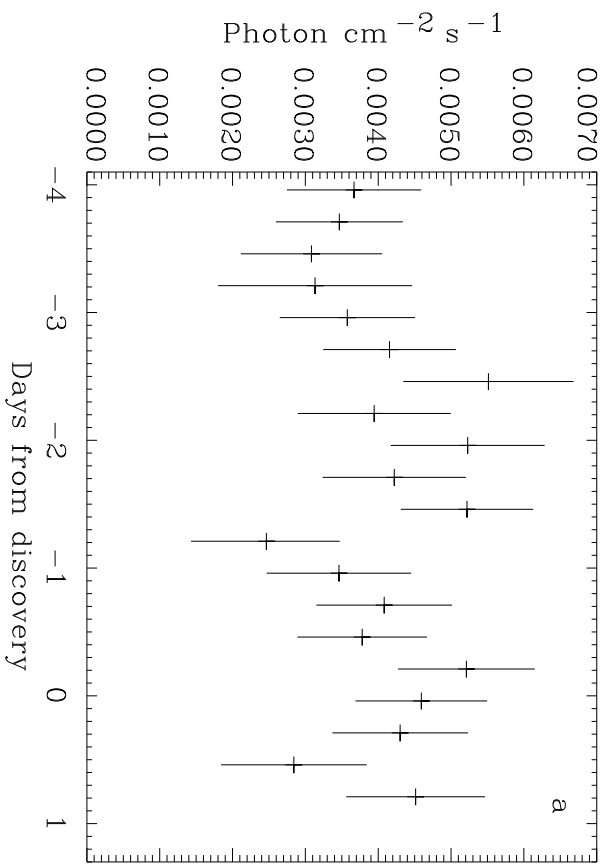
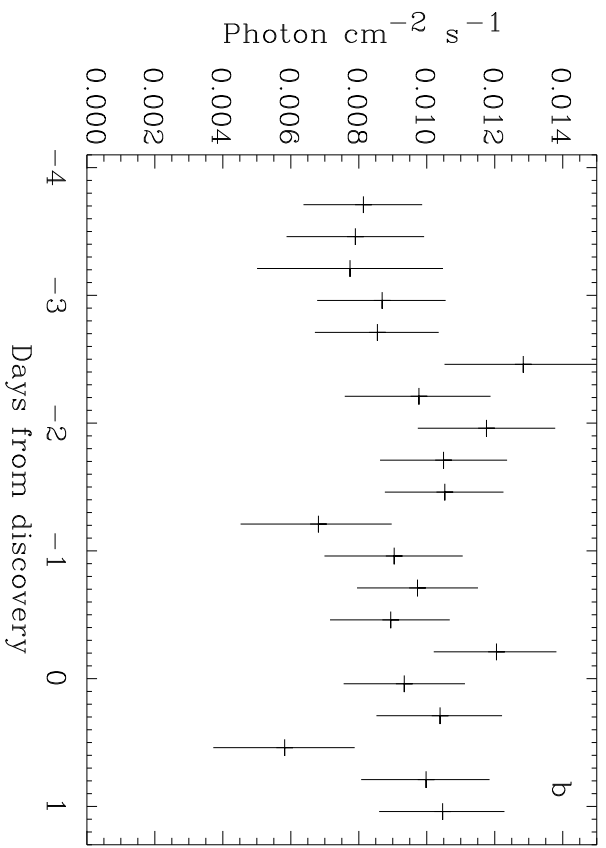
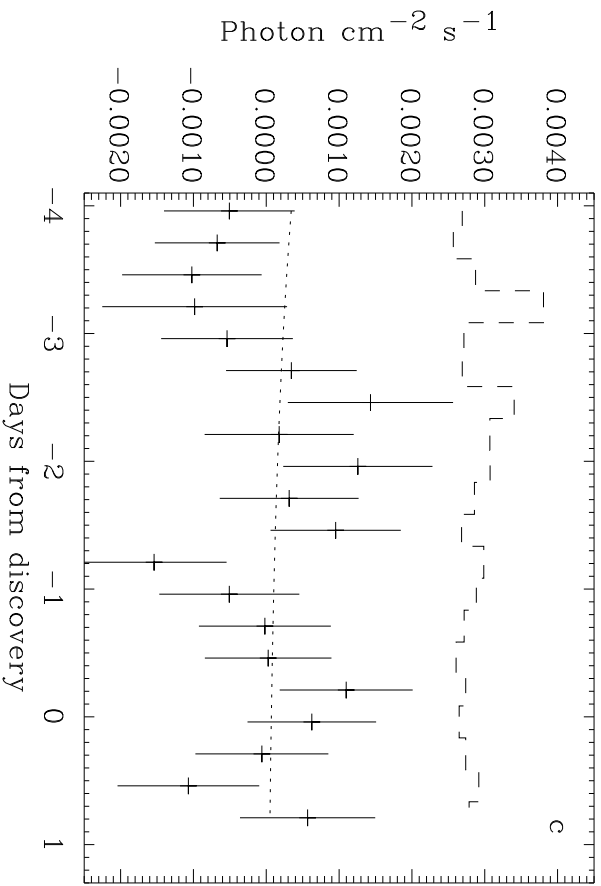
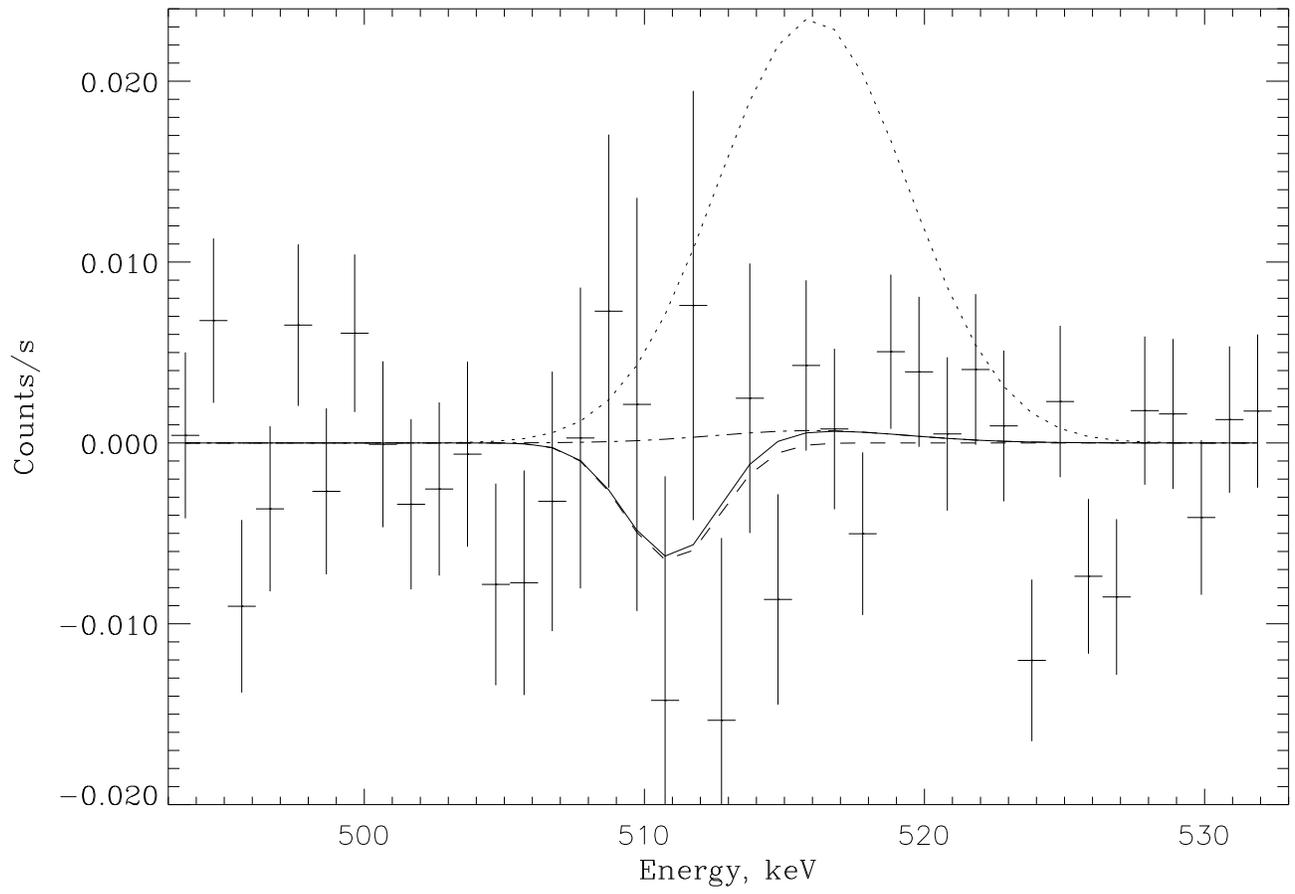
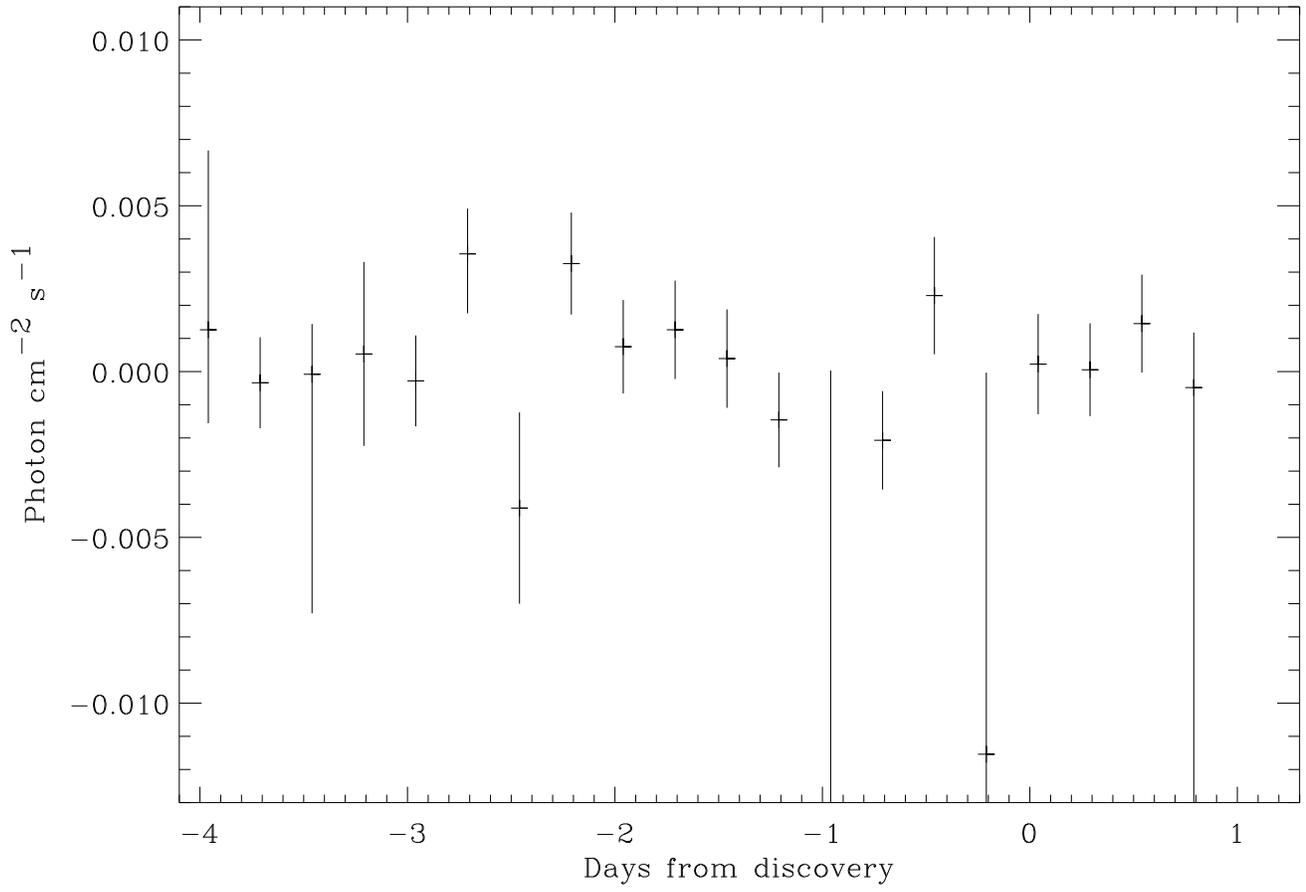
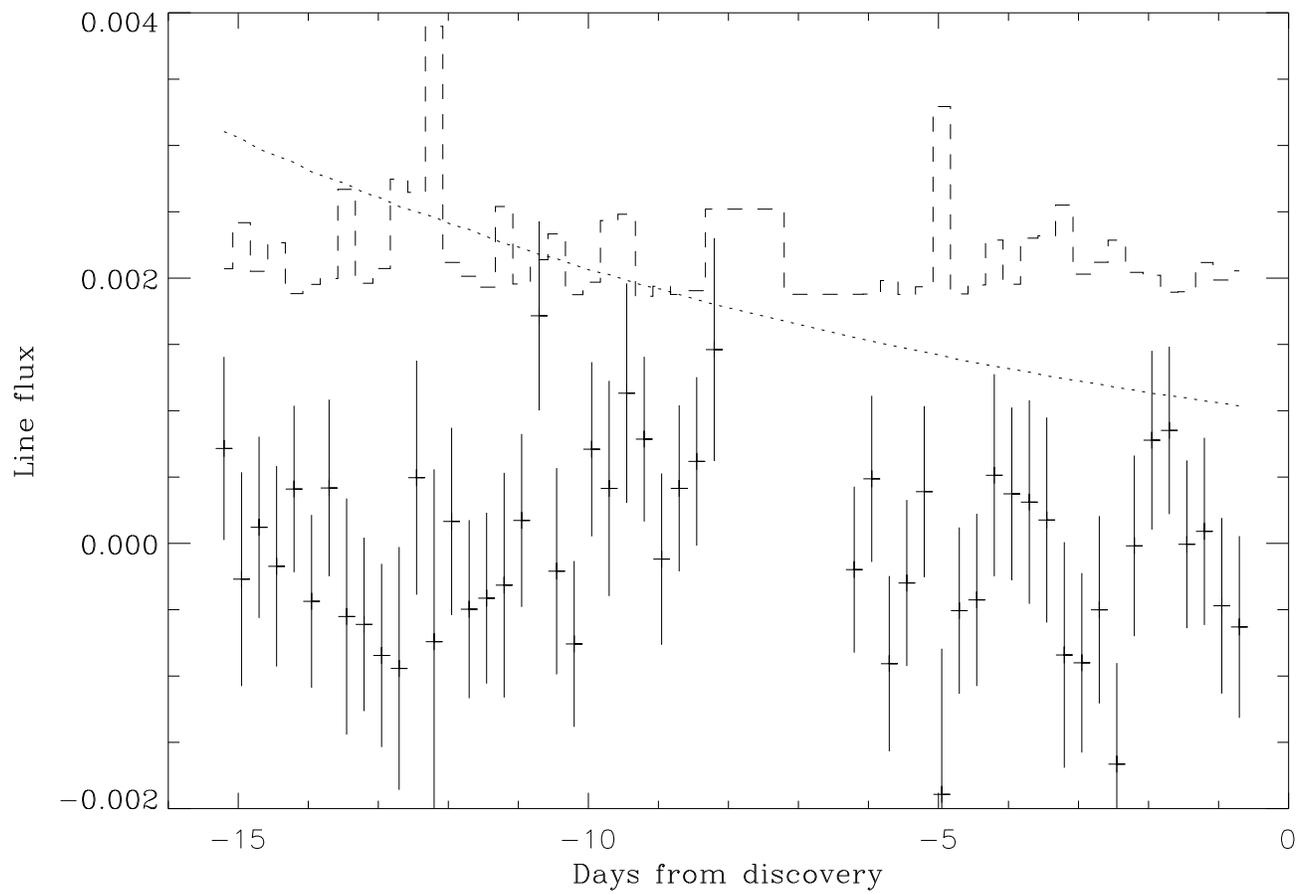
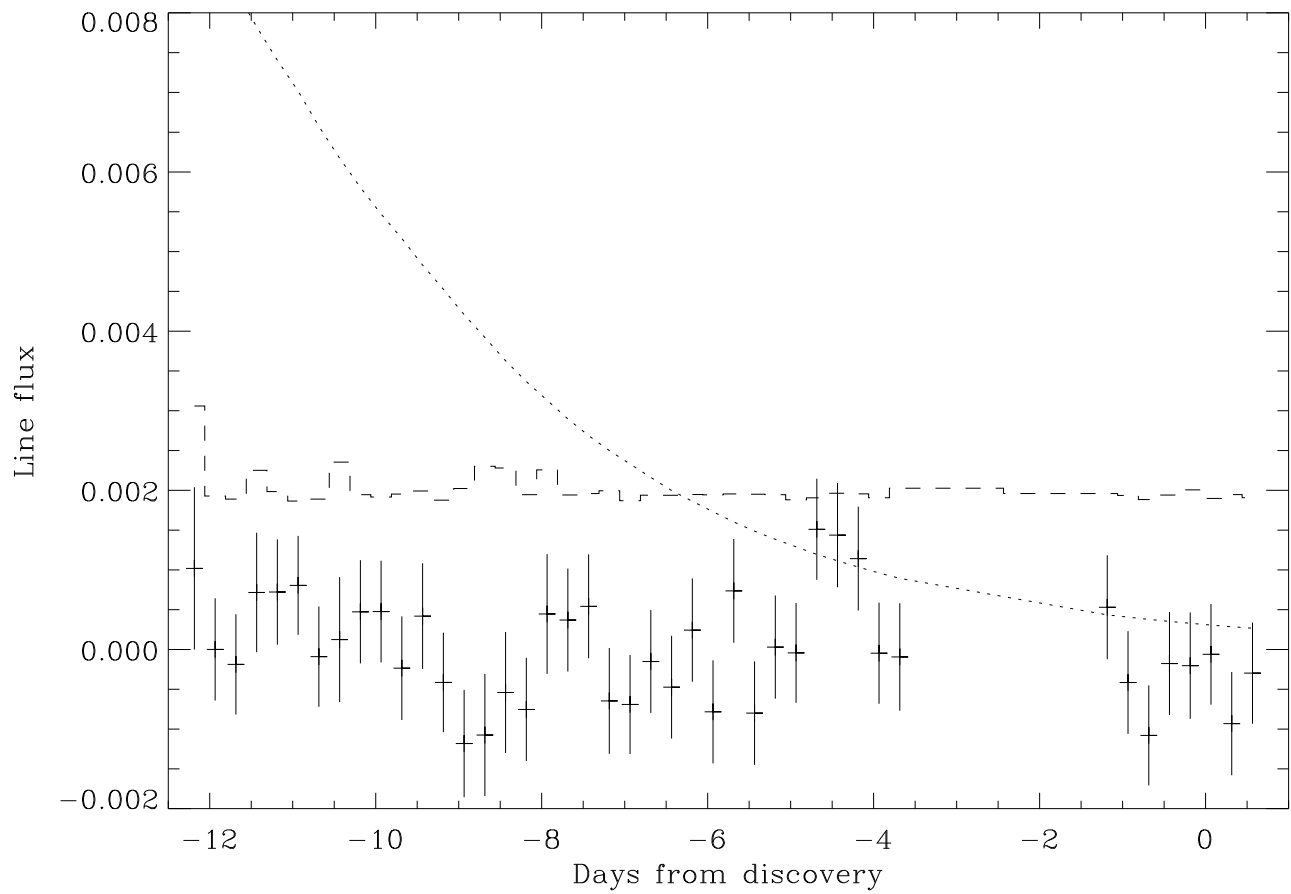
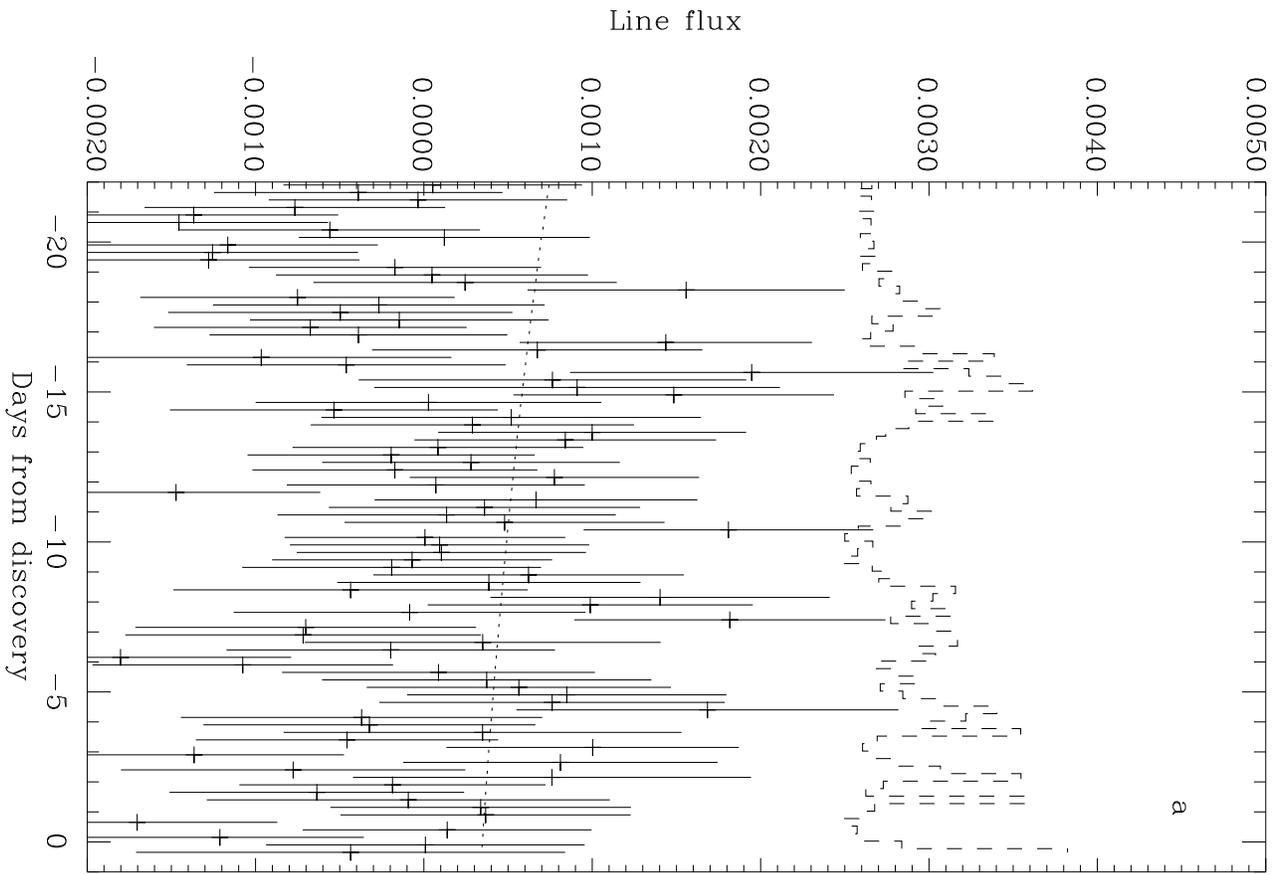
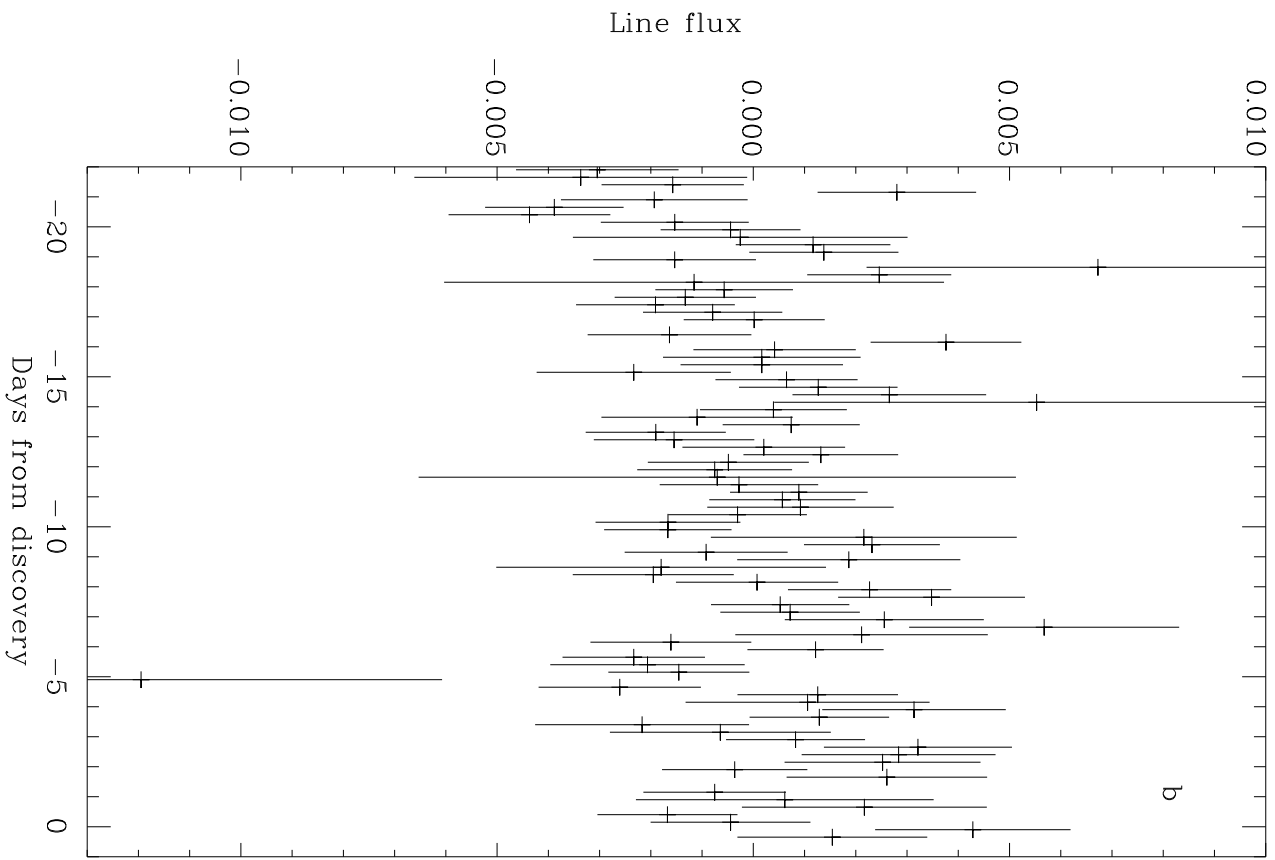
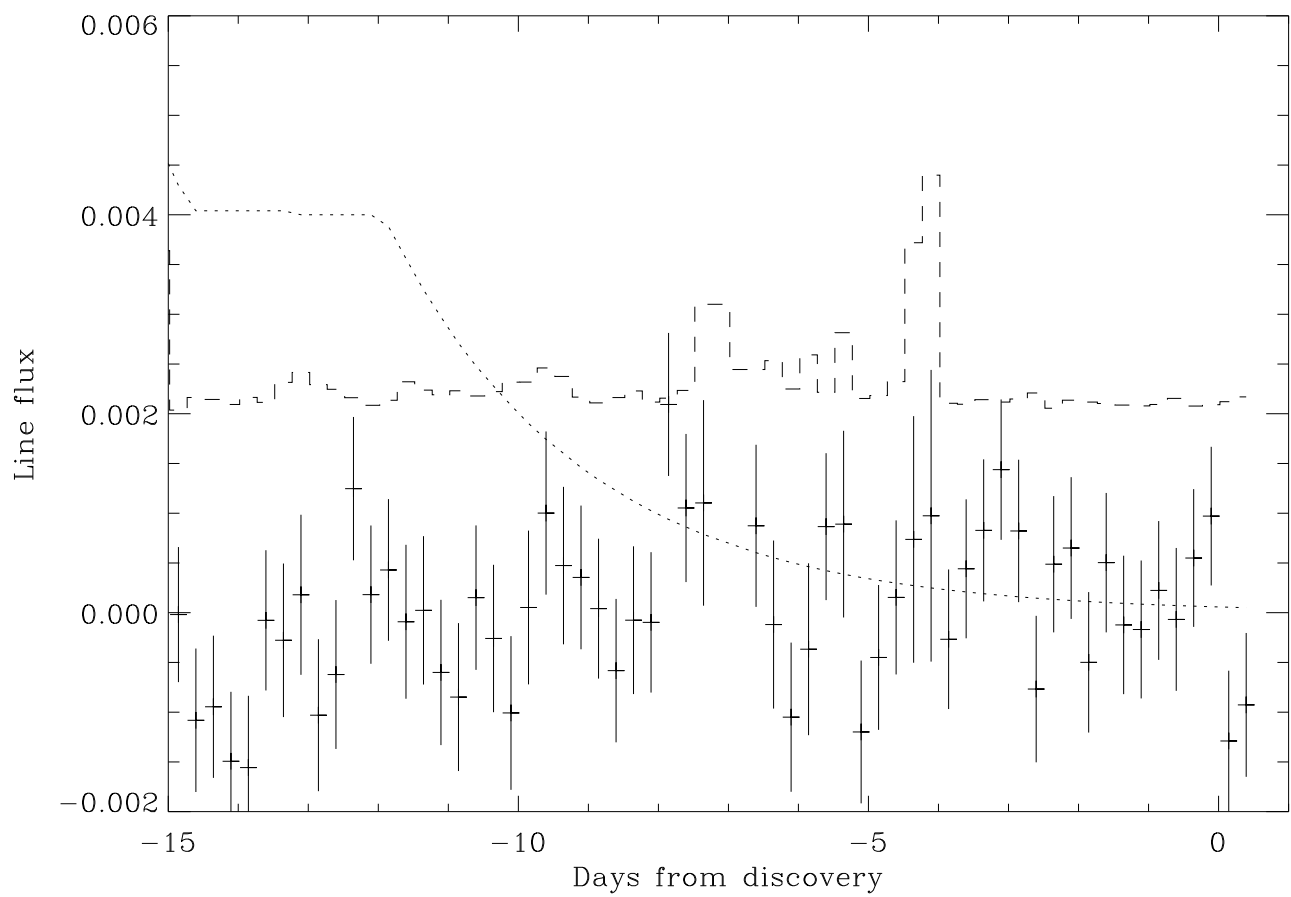
\end{document}